\documentclass[superscriptaddress,aps,prl,preprintnumbers,twocolumn,showpacs]{revtex4}

\usepackage{epsfig,amsmath,amssymb}
\usepackage[latin1]{inputenc}
\newcommand{\C}{{\cal{C}}}
\renewcommand{\epsilon}{\varepsilon}
\providecommand{\url}[1]{\texttt{#1}}

\begin{document}
%\preprint{{\bf DRAFT REVISED 7a}}

\title{Direct evaluation of large-deviation functions}

\author{Cristian Giardin\`a}
\affiliation{Eurandom, P.O. Box 513 -- 5600 MB Eindhoven, The
Netherlands}

\author{Jorge Kurchan}
\affiliation{PMMH-ESPCI, CNRS UMR 7636, 10 rue Vauquelin, 75005
  Paris, France}

\author{Luca Peliti}\thanks{Associato INFN, Sezione di Napoli.}
\affiliation{Dipartimento di Scienze Fisiche and Unità CNISM,
Università ``Federico~II'', 80126 Napoli, Italy}
\date{February 18, 2006}

\begin{abstract}
We introduce a numerical procedure to evaluate directly the
probabilities of large deviations of physical quantities, such
as current or density, that are local in time. The large-deviation functions
are given  in terms
of the typical properties of a modified dynamics, and since they
no longer involve
rare events, can be evaluated efficiently and over a wider ranges of values.
We illustrate the method with the current fluctuations
of the Totally Asymmetric Exclusion Process and with the
work distribution of a driven Lorentz gas.
\end{abstract}


\maketitle

In the last few years there has been a renewed interest in the
theory of large deviations of nonequilibrium systems, with the
development of general results concerning the fluctuations of soft
modes \cite{BDGJL2}, of nontrivial and rich analytic solutions of
explicit models \cite{BD,BD2,DL,DSt1,ED,ELS1,JV,schutz,spohn}, and
with the discovery of strikingly simple and general nonequilibrium
relations~\cite{EvCoMo,GaCo,Ja,Ku,LeSp,reviews,fourier,Ma,G-FDT}
(the Fluctuation Theorem, Jarzynski's relation) obeyed by work
fluctuations. Perhaps for the first time, we are gathering a few
glimpses of truly general features of macroscopic systems well out
of equilibrium.

The large-deviation function plays an essential role in the
investigation of nonequilibrium systems---a role akin to the free
energy in equilibrium ones. When available techniques do not allow
for an exact evaluation of this function, one turns to
simulations: but direct numerical simulation of large deviations
is hard, since, by definition, they are rare. In equilibrium, this
difficulty is often overcome by introducing biased (non-Boltzmann)
sampling~\cite{umbrella}. Here we show that a similar strategy can
be introduced in systems out of equilibrium, in order to evaluate
the large deviations function for quantities that are local in
time, although not necessarily in space. We find that, in
nonequilibrium, it is necessary not only to bias suitably the
dynamics of the system, but also to introduce a process by which
images (clones) of the system reproduce or die, a technique
inspired by the Diffusion Monte Carlo method~\cite{clones} of
quantum mechanics. In the present work, after deriving the general
formalism, we show its effectiveness by applying it to two
nonequilibrium processes: a stochastic one, the Totally Asymmetric
Exclusion Process (TASEP), and a deterministic one, a driven
Lorentz gas. Our algorithm allows to compute the probability of
obtaining a temporary large deviation  (compared to the typical) 
value  of the current in the first example and of the dissipated
work in the second one.

We consider the general setup of a system evolving according to
Markovian dynamics. Let $\C,\C'$  denote two configurations in the
 and let $U_{\C'\C}$ be the transition matrix
of the discrete (eventually continuous) time dynamics. Denoting by
$P_{\C}(t)$ the probability of being in the configuration $\C$ at
time $t$, one has
\begin{equation}
P_{\C'}(t+1)= \sum_{\C} U_{\C'\C}P_{\C}(t) \,.
\end{equation}
In a time interval of length $T$, a path $\C_0,\C_1,\ldots,\C_T$
in the configuration space, starting from a fixed 
$\C_0$, will have the probability
\begin{equation}
\mathop{\mathrm{Prob}}[\C_0,\C_1,\ldots,\C_T] = U_{\C_T
\C_{T-1}}\cdots\, U_{\C_2\C_1} \cdot U_{\C_1\C_0} \,.
\end{equation}
We shall consider physical quantities $Q_T$ that are additive in
time, i.e., which can be written as $Q_T= \sum_{t=0}^{T-1}
J(\C_{t+1},\C_t)$. For example, for a transition $\C \rightarrow
\C'$ in a lattice system:
\begin{equation}
J_{\C'\C}=\left\{ \begin{array}{ll}
1, & \textrm{ a particle jumps to the right;}\\
0, & \textrm{ nothing happens;} \\
-1,&  \textrm{ a particle jumps to the left.}
\end{array}\right.
\end{equation}
We are interested in calculating the probability of having a
current $Q_T$ in the time interval $T$, i.e., a current $q=Q_T/T$
per unit time. Denoting with angular brackets the average over trajectories,
we have
\begin{eqnarray}
\label{integral}
P\left(\frac{Q_T}{T} = q\right) &=& \left\langle
\delta(J_{\C_1\C_0}+\cdots\, +J_{\C_{T}\C_T-1} - q T)\right\rangle \nonumber \\
&=& \frac{1}{2\pi i}
\int_{-i\infty}^{+i\infty} d\lambda \,e^{T \,[\,\mu(\lambda)- \lambda q\, ]},
\end{eqnarray}
where we used the integral representation of delta function and we have
defined
\begin{eqnarray}
e^{T \mu(\lambda)} &=& \left\langle e^{\lambda
(J_{\C_1\C_0}+\cdots \,+J_{\C_{T}\C_{T-1}})} \right\rangle \\
=& & \sum_{\C_1,\ldots,\C_T} U_{\C_T\C_{T-1}}\cdots\, U_{\C_1\C_0}
e^{\lambda (J_{\C_1\C_0}+\cdots\, +J_{\C_{T}\C_{T-1}})}.\nonumber
\end{eqnarray}
In the limit $T\rightarrow\infty$, by applying the steepest
descent method to Eq.~(4) (and assuming that the imaginary contour
line can be deformed to the real line), one obtains that the large
deviation function $f(q) = \lim_{T\rightarrow \infty}\ln P(Q_T)/T$
and $\mu(\lambda)$ are Legendre transforms of each other:
\begin{equation}
f(q) = \max_{\lambda} [\mu(\lambda) - \lambda q],
\end{equation}
so that $q=\mu'(\lambda^*)$ where $\lambda^*$ is the saddle.
We introduce the bias (\ref{paths}) of the original measure
by defining the new matrix
\begin{equation}
\label{paths}
\tilde U_{\C'\C} \equiv e^{\lambda J_{\C'\C}} U_{\C'\C},
\end{equation}
so that
\begin{equation}
e^{T \mu(\lambda)} =
\sum_{\C_1,\ldots,\C_T} \tilde U_{\C_T\C_{T-1}}\cdots\, \tilde U_{\C_1\C_0} =
\sum_{\C_T} \left[ \tilde U^{T}\right]_{\C_T\C_0}.
\label{R}
\end{equation}
Introducing the spectral decomposition
$\tilde U = \sum_{j} e^{\Lambda_j} \;|\Lambda^R_j\rangle \langle \Lambda^L_j |$
where we assumed a complete biorthogonal set of the matrix $\tilde U$ exists, i.e.
$\tilde U |\Lambda^R_j\rangle  = e^{\Lambda_j} |\Lambda^R_j\rangle$
$ \langle \Lambda^L_j|\tilde U = e^{\Lambda_j} \langle \Lambda^L_j|$
and denoting by $e^{\Lambda}$ the eigenvalue of $\tilde U$
with the largest real part, and by $|\Lambda^R\rangle$,
$|\Lambda^L \rangle$ the corresponding right and left
eigenvectors, we have, for large times $T$,
\begin{equation}
e^{T \mu(\lambda)} = \sum_{\C_T} \langle \C_T|\Lambda^R\rangle \langle \Lambda^L|\C_0
\rangle \; e^{T \Lambda},
\label{sp1}
\end{equation}
so that $\Lambda=\mu(\lambda)$.
In order to compute $\mu(\lambda)$, one possibility is to perform
path-sampling over the trajectories with weight (\ref{paths}).
Such a procedure has been proposed in the context of the work
distributions~\cite{biased} on nonequilibrium trajectories. In
this paper we propose a different strategy: the idea is to define
a new effective dynamics whose expectation values directly give
the large deviations. As we shall see, the new dynamics involves
the parallel evolution of clones which reproduce and die, a
procedure inspired by the ``Diffusion Monte Carlo'' method of
simulation of the Schrödinger equation~\cite{clones}. In order to
write eq.~(\ref{R}) as expectation on the new dynamics, let us put
$K_\C \equiv \sum_{\C'} \tilde U_{\C'\C}$, and define the
stochastic matrix
\begin{equation}
U_{\C'\C}' \equiv  \tilde U_{\C'\C}  K_{\C}^{-1}.
\end{equation}
We now have, instead of eq.~(\ref{sp1}),
\begin{equation}
e^{T\mu(\lambda)} =
\sum_{\C_2,\ldots,\C_T}  U_{\C_T\C_{T-1}}' K_{\C_{T-1}}
\cdots  U_{\C_1\C_0}' K_{\C_0}.
\end{equation}
This can be realized by considering an {\em ensemble} of $L$
copies (``clones'') of the system, and by successively going, for
all of them, through a process defined by the following three
steps:
\begin{itemize}
\item A cloning step:
\begin{equation}
P_{\C}(t+1/2)= K_{\C} P_{\C}(t),
\end{equation}
where the configuration $\C$ of the selected copy gives rise to
$G$ identical clones, $G=[K_{\C}]+1$ with probability
$K_{\C}-[K_{\C}]$, and $G=[K_{\C}]$ otherwise ($[x]$ denotes the
integer part of $x$). If $[K_{\C}]=0$, the copy may be killed and
leave no offspring.

\item  A shift step without cloning of all the
offspring of $\C'$ with the modified dynamics $U'$
\begin{equation}
P_{\C'}(t+1) = \sum_{\C} U'_{\C'\C} P_{\C}(t+1/2) .
\end{equation}

\item An overall cloning step with an adjustable rate $M_t =
L/(L+G)$ (at each time the same for all configurations), so as to
keep  the total number of clones constant. This amounts to
multiplying $\tilde U$ by $M_t$ times an identity, at each time.
\end{itemize}

It is easy to see that, in the long-time limit, the compensatory
factor gives us $\mu(\lambda)$ through
\begin{equation}
- \ln [M_T \cdots M_2 \cdot M_1] = T \mu(\lambda).
\label{mumu}
\end{equation}

{\em Remark 1:} We note that, if the quantity $Q_T$, whose 
deviations we wish to compute, depends on a single configuration
rather than on a pair of configurations, such as for the density
$Q_T = \sum_{t=1}^T \rho(\C_t)$, the same derivation goes through
with the substitution $J_{\C'\C} \rightarrow \rho(\C)$.

{\em Remark 2:} The configurations obtained in the course of the 
simulation are representative of the typical ones {\em at the end}
($t=T$),
{\em rather than within} ($0 \ll t \ll T$) the interval of time $T$ 
during which the large deviations are observed. (Their probabilities
 are proportional to$ \langle \C|\Lambda^R \rangle$ and 
$ \langle \Lambda^L|\C \rangle \langle \C|\Lambda^R \rangle$,
respectively).

%{\em Remark 2:} One should not consider the configurations
%obtained in the course of the simulation as representative of the
%typical configurations of the system in the time evolution
%conditioned by the imposed values of the large deviation parameter
%$\lambda$. Indeed, the time evolution of the system up to $T$ time
%steps does not depend on the values of the current at later times,
%whereas the weight on the configuration depends on the whole
%history of the system. Thus the probability that a configuration
%$\C$ is visited after $T$ steps of the dynamics described above is
%$\sim  \langle \C|\Lambda^R \rangle$, while the weight of the same
%configuration at time $T$, for a system evolving for a much longer
%time, is given by $\sim \langle \Lambda^L|\C \rangle \langle
%\C|\Lambda^R \rangle$. As we shall see, they contain however
%interesting information.

%{\em Remark 2:} The probability that a configuration $\C$ is
%visited after $T$ steps of the dynamics described above is $\sim
%\sum_{\C'}\langle \C'|\C \rangle \langle \C|\Lambda^R \rangle$:
%this corresponds to the typical configurations \textit{at the end}
%of the time-interval in which we are conditioning the large
%deviation, in general different from the typical configurations
%{\em well inside} this interval (which  are visited with
%probability $\sim \langle \Lambda^L|\C \rangle \langle
%\C|\Lambda^R \rangle$). As we shall see, they contain however
%interesting information.

We now turn to two examples: the Totally Asymmetric Exclusion
Process, and the Lorenz gas.

\paragraph*{A stochastic system: The Totally Asymmetric Exclusion
Process (TASEP).}

The TASEP \cite{spohn} consists of particles on a ring with discrete sites with
occupancy zero or one. A given particle chosen at random does not
attempt to move with probability $(1-\alpha)$, and with
probability $\alpha$ attempts to move to the right and succeeds
if  the corresponding site is empty. The parameter $\alpha$ can be
made small to approach the continuous time limit. Here we shall
set it to unity. Let us denote by $X_ {\C}$ the number of
different configurations that can be reached by making a
one-particle move (1PM) from $\C$. Then the non-zero entries of
$U_{\C'\C}$ are given by
\begin{equation}
U_{\C'\C}=\left\{ \begin{array}{ll}
\alpha/N, & \textrm{if } \C \rightarrow \C'
\textrm{ is a 1PM;} \\
1- (X_ {\C}\alpha/N), & \textrm{if } \C'=\C.
\end{array}\right.
\end{equation}
This implies for $\tilde U$
\begin{equation}
\tilde U_{\C'\C}=\left\{ \begin{array}{ll}
\alpha e^{\lambda}/N, &
 \textrm{if } \C \rightarrow \C'
\textrm{ is a 1PM;}\\
1- (X_{\C} \alpha/N) & \textrm{if } {\C'} ={\C}.
\end{array}\right.
\end{equation}
Thus, for a configuration ${\C}$ with $X_{\C}$ mobile particles,
we have
\begin{equation}
K_{{\C}} = 1 + \frac{X_{\C}\alpha}{N}(e^{\lambda} -1),
\end{equation}
and finally
\begin{equation}
U_{\C'\C}'=\left\{ \begin{array}{ll}
(\alpha e^{\lambda}/N)/K_{\C}, &
 \textrm{if } \C \rightarrow \C'
\textrm{ is a 1PM,}\\
(1- X_{\C'} \alpha/N)/K_{\C}, & \textrm{if } \C'=\C.
\end{array}\right.
\end{equation}
Thus, with probability $(1- X_{\C} \alpha/N)/K_{\C}$ no move
is made; otherwise we move a particle randomly chosen with uniform probability
among the $X_{\C}$ mobile particles.

\begin{figure}[ht]
%\begin{center}
\includegraphics[width=6cm]{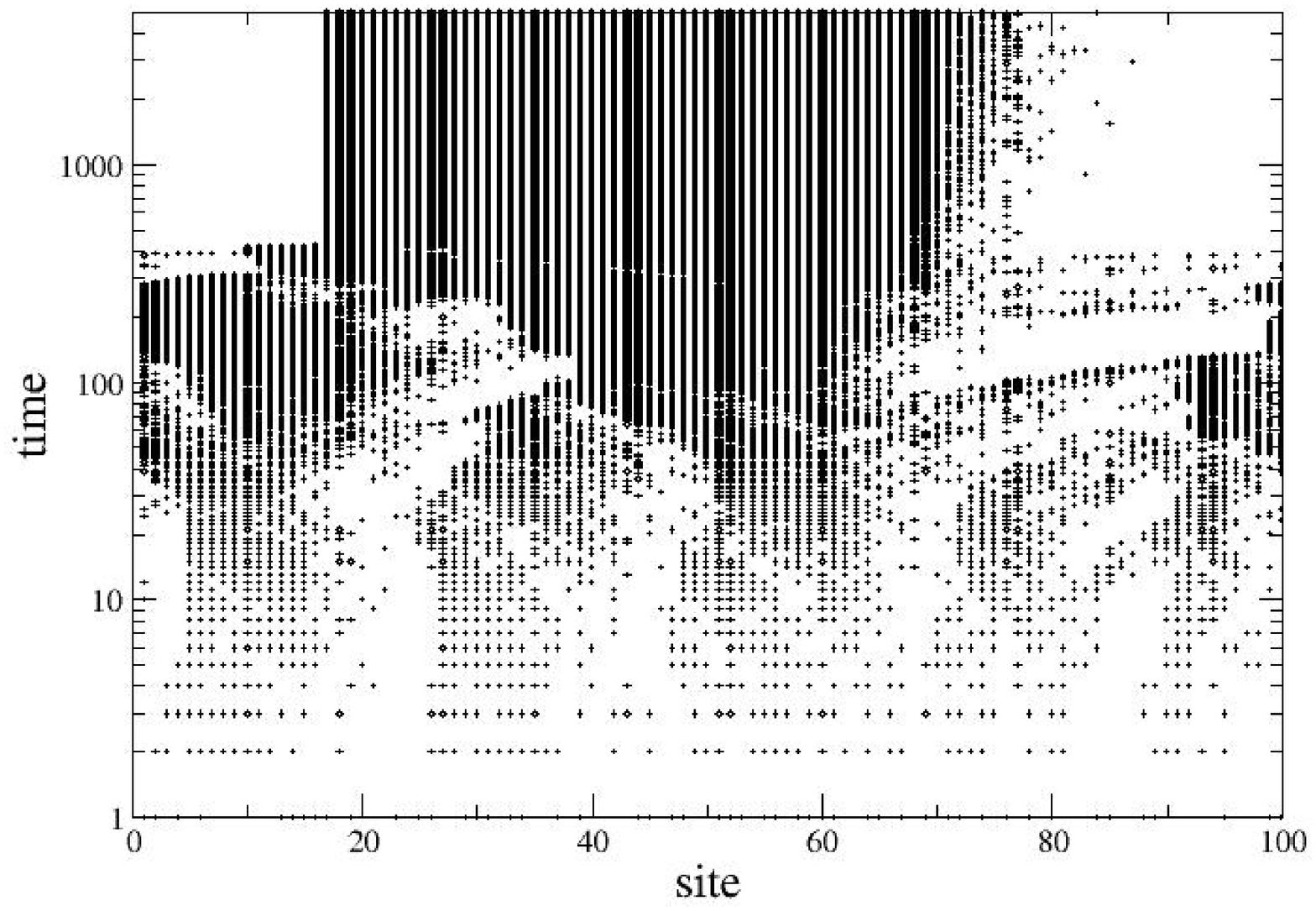}
\includegraphics[width=6cm]{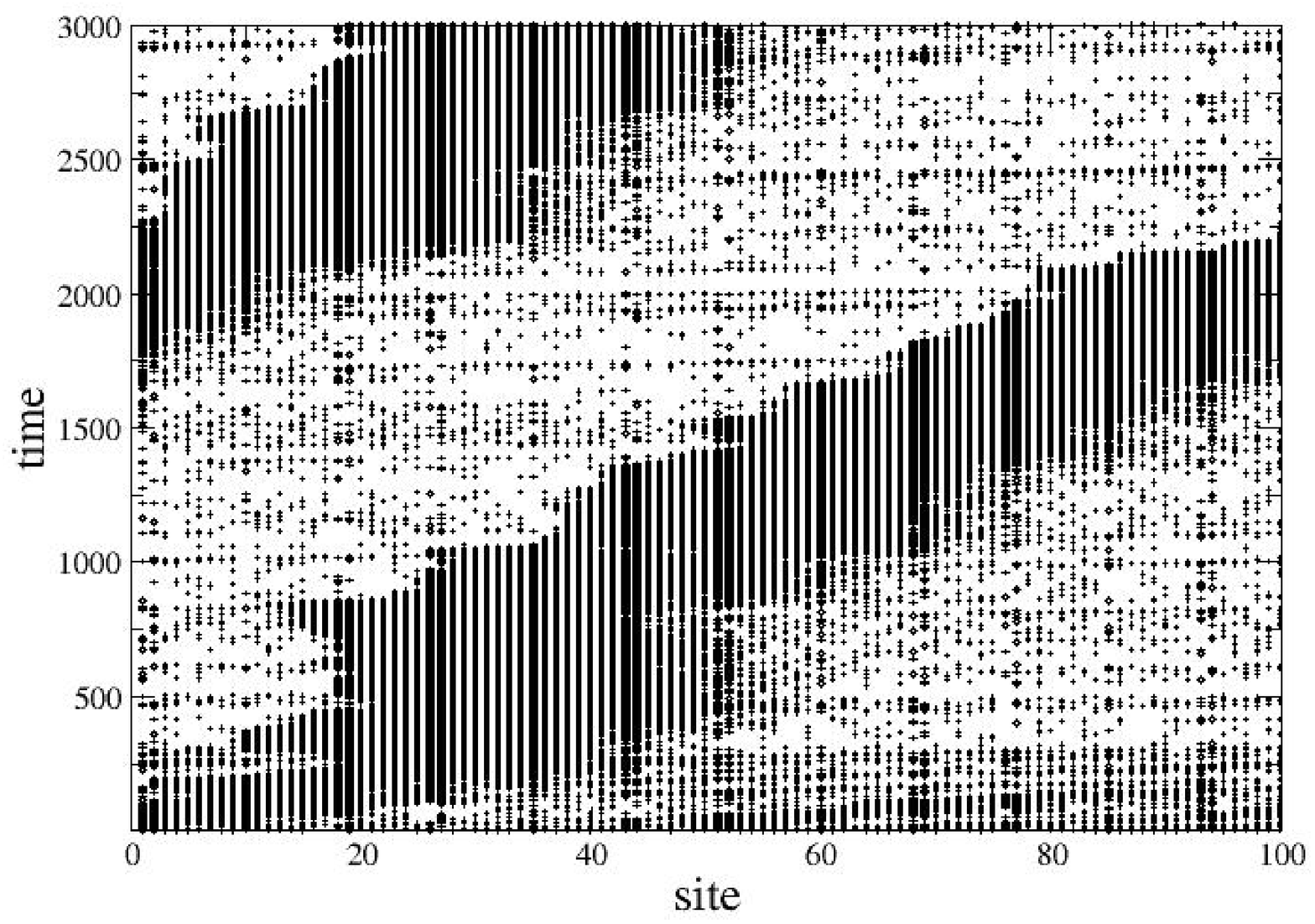}
%\end{center}
\caption{Space-time diagram for a ring of $N=100$ sites.
Top: $\lambda=-50$ and density $0.5$; the
 shock is dense and does not advance. 
Bottom: $\lambda=-30$ and density $0.3$;
the shock drifts to the right.}
\label{cinco}
\end{figure}

\begin{figure}[hb]
\begin{center}
\includegraphics[width=6.cm]{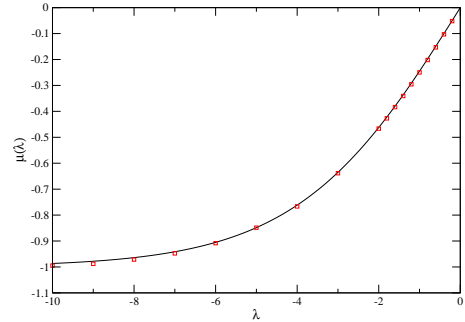}
\end{center}
\caption{Plot of $\mu(\lambda)$ vs.~$\lambda$ for the TASEP at
density one-half. Numerical results and analytic results of
ref.~\cite{BD2}, with points and full line, respectively.}
\label{dos}
\end{figure}

In Figure \ref{cinco} (top) we show a space-time diagram of the system
with $N=100$ particles, density $0.5$ and $\lambda=-50$. The
simulation was done with $L=1000$ clones, each of them initialized
with random (uniform) occupancy numbers, such that the
configuration has density $0.5$. We notice that the configurations
rapidly become inhomogeneous, exhibiting an alternation of a
regions with high density with regions of slow density, as in
traffic jams or in shock waves. The high-density regions
eventually coalesce into a single one. The figure does not quite
represent the evolution of a shock (because of Remark~2 above),
but rather the configuration at the end of the time-interval for
time intervals ending at progressively longer times. As predicted
by the theory for this value of the density, the shock does not
drift, although different initial conditions lead to different
shock positions. Bottom of Figure \ref{cinco} shows the case $\lambda=-30$,
and density $0.3$: we see that the shock has a net drift to the
right, again as predicted by the theory~\cite{BD2}. Finally, in
figure \ref{dos} we show the numerical results obtained for
$\mu(\lambda)$, and compare them to the analytic ones of
Ref.~\cite{BD2}. The agreement is excellent, and the numerical
effort corresponds to tens of minutes of a personal computer time.

%
%\begin{figure}[ht]
%\begin{center}
%\includegraphics[width=8.cm]{confi-lambda30.eps}
%\end{center}
%\caption{Space-time diagram for a ring of $N=100$ sites,
%$\lambda=-30$ and density $0.3$.
%The shock drifts to the right.}
%\label{uno}
%\end{figure}
%

%

\paragraph*{A deterministic system: The Lorentz Gas and the
Gallavotti-Cohen theorem.}

\begin{figure}
\includegraphics[width=5.cm]{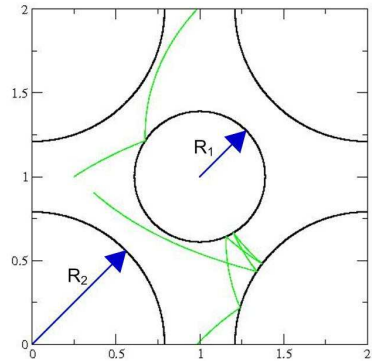}
\caption{The billiard. The radii are $R_1 = 0.39, R_2=0.79$. We
also show an example of trajectory for the external field $\vec
E=(1,0)$.} 
\label{tres}
\end{figure}

This system consists of a number of particles (in our case only
one) moving inside a billiard as in figure \ref{tres}, with
periodic boundary conditions. The particle is under the action of
a force field $\vec E$, and is subject to a deterministic
thermostat that keeps the velocity modulus constant $|\vec v|=1$.
Between bounces, the equations of motion are:
\begin{eqnarray}
\ddot x_i &=& - E_i +\gamma(t) \dot x_i ,\qquad i=1,2;\nonumber \\
\gamma(t) &=& \sum_i E_i \dot x_i.
\end{eqnarray}
We wish to compute the generating function of the dissipated work
$Q_T = \int_0^T \gamma(t)\, dt$, and check the Gallavotti-Cohen
theorem, which states that 
$P(Q_T)/P(-Q_T) = \exp(Q_T)$, which is equivalent, thanks to 
Eq. (\ref{integral}), to the symmetry of $\mu(\lambda)$ around
$\lambda=-\frac{1}{2}$.

The dynamics is deterministic, and hence cloned systems will
evolve together and perform a poor sampling. To get around this
problem, we introduce a small stochastic noise, and check the
stability of results in the limit of small noise. We evolve the
system for \textit{macroscopic} intervals ${\cal{T}}$, and clone
with a factor $K_t=e^{\lambda \Gamma_{\cal{T}}(t)}$, where
$\Gamma_{\cal{T}}(t) = \int_t^{t+{\cal{T}}}\gamma(t) \,dt $ is the
total dissipated work over the interval. Before each
deterministic step of time ${\cal{T}}$, clones are given random
kicks of variance $\Delta$ in position  and/or velocity direction.
\textit{The time-interval ${\cal{T}}$ and the noise intensity
$\Delta$ are chosen so that twin clones have a chance to separate
during time ${\cal{T}}$}, and this depends on the chaotic
properties of the system. In the present case, we checked that
$0.1\leq {\cal{T}}\leq 1$  allows for a few collisions, which
guarantees clone diversity for $10^{-3}\leq \Delta \leq 10^{-4}$.

\begin{figure}
\begin{center}
\includegraphics[width=6.cm,angle=-90]{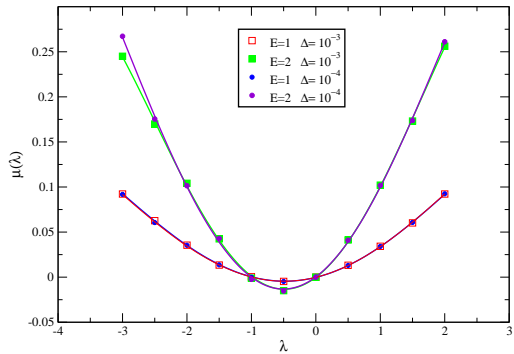}
\end{center}
\caption{Plot of $\mu(\lambda)$ vs.\ $\lambda$ for the driven
Lorentz gas. Data for $\vec E=(E,0)$, $E=1,2$ and noise intensity
$\Delta = 10^{-3},10^{-4}$. The Gallavotti-Cohen theorem implies
the symmetry around $\lambda=-1/2$. The best fit (continuous lines)
is quadratic for $E=1$ (gaussian
behavior), and a 4-th order polynomial for $E=2$. 
%Inset:
%The billiard with   $R_1 = 0.39, R_2=0.79$, and 
% a trajectory for  $\vec
%E=(1,0)$.
} \label{cuatro}.
\end{figure}

In Fig.~\ref{cuatro} we show the results of $\mu(\lambda)$
%(related to the average entropy production per unit time $q$ by
%$q=\mu'(\lambda)$) 
for $-3 \leq \lambda \leq 2$, and for $\vec
E=(E,0)$ with $E=1,2$, corresponding to very large current
deviations.

Diversity in a population of reproducing units is maintained by a
balance between the natural loss due to sampling fluctuations and
the increase introduced by mutations, represented in our case by
noise~\cite{Kimura}. Thus, if the noise level is too small in the
billiard case, most of the clones correspond to too close
configurations, and our results become noisy and unreliable. The
same phenomenon explains why all clones exhibit shocks in
essentially the same position for any given run in the TASEP
(since they share a common ancestor): but we found that in this
case the phenomenon poses no problem for the sampling, since the
current does not depend on the position of the shock.

In conclusion, we have shown that sampling methods based on a
modified dynamics with clones can be used to efficiently compute
the large deviations function, in  times and within ranges of
values that cannot be reached in a direct simulation.

\begin{acknowledgments}
We wish to thank B. Derrida for his encouragement and suggestions,
and S. Tanase-Nicola for making us aware of Refs.~\cite{biased}.
CG thanks ESPCI for kind hospitality and acknowledges NWO-project
613000435 for financial support. LP thanks the LPTMS, Universit\'e
Paris-Sud, for hospitality.
\end{acknowledgments}


\begin{thebibliography}{}

\bibitem{BDGJL2} L.  Bertini, A.  De Sole, D.  Gabrielli, G.
Jona-Lasinio, C.  Landim,
J. Stat. Phys. {\bf 107},  635-675  (2002);
Phys. Rev. Lett. {\bf 94},  030601  (2005);
 cond-mat/0506664 (2005).

\bibitem{BD}
 T. Bodineau, B. Derrida,
 Phys. Rev. Lett. {\bf 92}, 180601 (2004).
 \bibitem{BD2}
T. Bodineau, B. Derrida, cond-mat/0506540.

\bibitem{DL}
 B. Derrida, J.L. Lebowitz,  Phys. Rev. Lett. {\bf 80}, 209-213 (1998).

\bibitem{DSt1}
M. Depken, R. Stinchcombe, Phys. Rev. Lett. {\bf 93}, 040602 (2004);
Phys. Rev. E {\bf 71}, 036120 (2005).

\bibitem{ED}
 C. Enaud, B. Derrida, J. Stat. Phys. {\bf 114}, 537 (2004).

\bibitem{ELS1}
G.~Eyink, J.L.~Lebowitz, H.~Spohn,
Comm. Math. Phys. {\bf 132},  253  (1990).

\bibitem{JV}
L. Jensen, Ph.D. dissertation, New York University, (2000);
C. Landim, S. Olla, S.R.S.Varadhan,  in:
{\it Stochastic Analysis on Large Interacting Systems},
(T. Funaki, H. Osada eds.)
Adv. Stud. Pure Math., {\bf 39}, 1 (Mathematical Society of Japan, Tokyo, 2004).

\bibitem{schutz}
G.M. Sch\"utz,
in: {\it Phase Transitions and Critical Phenomena}, (C. Domb und J. Lebowitz (eds.))
{\bf 19}, 1 (Academic Press, London, 2000).

 \bibitem{spohn} H.~Spohn, \textit{Large Scale Dynamics of Interacting
Particles}, Springer-Verlag, Berlin, 1991;
C. Kipnis, C. Landim, \textit{Scaling Limits of Interacting Particle Systems},
Springer-Verlag, Berlin, New York, 1999.


\bibitem{EvCoMo}
D. J. Evans, E. G. D. Cohen, and G. P. Morriss, Phys. Rev. Lett. {\bf 71},
2401 (1993);
 D. J. Evans and D. J. Searles, Phys. Rev. E {\bf 50}, 1645 (1994).

\bibitem{GaCo}
G. Gallavotti and E.G.D. Cohen, Phys. Rev. Lett. {\bf 74}, 2694 (1995);
 J. Stat. Phys. {\bf 80}, 931 (1995).

\bibitem{Ja}
C. Jarzynski, Phys. Rev. Lett. {\bf 78}, 2690 (1997);
 Phys. Rev. E {\bf 56}, 5018 (1997).

\bibitem{Ku} J. Kurchan, J. Phys. A (Math. Gen.) {\bf 31},
3719 (1998).

\bibitem{LeSp} J. L. Lebowitz and H. Spohn, J. Stat. Phys. {\bf 95}, 333 (1999).

\bibitem{reviews} See
D. Evans and D. Searles, Adv. Phys. {\bf 51}, 1529 (2002);
and  the corresponding  chapter of
G. Gallavotti,  \textit{Statistical Mechanics:
A Short Treatise},
Texts and Monographs in Physics, Springer Verlag
(1999), \url{http://ipparco.roma1.infn.it/pagine/libri.html}.




\bibitem{fourier}     J.P.   Eckmann, C.A. Pillet, L. Rey-Bellet,
J. Stat. Phys.  {\bf 95}, 305 (1999); L.  Rey-Bellet, L.E. Thomas,
Ann. Henri Poincaré {\bf 3}, 483 (2002).


\bibitem{Ma}C. Maes, J. Stat. Phys. {\bf 95}, 367 (1999).

\bibitem{G-FDT} G.  Gallavotti, Phys. Rev. Lett. {\bf 77} 4334 (1996).

\bibitem{umbrella}J. P. Valleau and D. N. Card, J. Chem. Phys. {\bf 57}
5457 (1972); G. M. Torrie and J. P. Valleau, J. Comp. Phys. {\bf
23} 187, (1977).


\bibitem{clones} Cloning in the simulation
of Schrödinger's equation was introduced in: J. B. Anderson, J.
Chem. Phys. {\bf 63}, 1499 (1975). For a discussion of error
sources and references, see: N. Cerf and O. C. Martin, Phys. Rev.
E {\bf 51}, 3679 (1995). See also the recent:  M. Rousset and G. Stoltz
cond-mat/0511412.

\bibitem{biased} See, e.g., S. X. Sun, J. Chem. Phys. {\bf 118}, 5769 (2003),
and H. Oberhofer, C. Dellago and P. L. Geissler,
J. Chem. Phys. B {\bf 109}, 6902 (2005),  and references therein.




\bibitem{Kimura}See, e.g., M. Kimura, \textsl{The Neutral Theory
of Molecular Evolution} (Cambridge: Cambridge U. P., 1983), in
particular Chap.~9.

\end{thebibliography}
\end{document}